\documentclass[11pt,twoside]{article}
\usepackage{asp2014}

\aspSuppressVolSlug
\resetcounters

\begin{document}

\title{Towards precise and accurate Cepheid chemical abundances\\ \centering for 1$\%$ $\mbox{H}_0$ measurement: temperature determination}
\author{Sara~Mancino,$^{1,2,3}$ Martino~Romaniello,$^1$ Richard~I.~Anderson,$^1$ and Rolf-Peter~Kudritzki$^{2,4}$}

\affil{$^1$European Southern Observatory, Garching b. M\"{u}nchen, Germany}
\affil{$^2$Ludwig~Maximilians~Universit\"{a}t, M\"{u}nchen, Germany;}
\affil{$^3$Excellence~Cluster~Universe, Garching b. M\"{u}nchen, Germany }
\affil{$^4$Institute for Astronomy, University of Hawaii at Manoa}

\email{smancino@eso.org}
\paperauthor{Sara~Mancino}{smancino@eso.org}{}{European Southern Observatory}{}{Garching b. M\"{u}nchen}{Bayern}{85748}{Germany}
\paperauthor{Martino~Romaniello}{mromaniello@eso.org}{}{European Southern Observatory}{}{Garching b. M\"{u}nchen}{}{85748}{Germany}
\paperauthor{Richard~I.~Anderson}{randerson@eso.org}{}{European Southern Observatory}{}{Garching b. M\"{u}nchen}{}{85748}{Germany}

\paperauthor{Rolf-Peter~Kudritzki}{kud@ifa.hawaii.edu}{}{Institute for Astronomy, University of Hawaii at Manoa}{}{Honolulu}{96822}{Hawaii}{USA}

\begin{abstract}

One of the outstanding problems in modern physics is the origin is of accelerated expansion of the universe. High-precision determinations of the Hubble parameter $\mbox{H}_0$ at different redshifts provide direct insight into the Universe expansion: equation of state of the Dark Energy, geometry and curvature of space, sum of neutrino masses and isotropy of the Universe. We investigate the effect of chemical composition on the classical distance ladder based on Cepheids stars and Supernovae type Ia. Cepheids belonging to Magellanic Clouds represent a natural anchor for the ladder, due to their proximity and the very well known geometric distance. 
Accordingly, the LACES collaboration collected the largest spectroscopic sample of MCs Cepheids, containing more than 300 stars and more than 1200 medium-high resolution spectra. Abundance measurements, performed by equivalent width and curve of growth analysis, have been tested on a wide grid of synthetic spectra to quantify the systematic arising from this procedure. We focus on the errors arising from temperature measurements, using the Line Depth Ratio method. We find that in order to not bias the final results many aspects (e.g. degeneracy of the atmospheric parameters, instrumental set up etc.) should be explicitly taken into account. In particular, there is a not negligible metallicity dependence.

\end{abstract}

\section{Rationale: $\mbox{H}_0$ late- and early-time measurements}

$\mbox{H}_0$ marks the present universe expansion rate, and its precise value helps to constrain the other cosmological parameters, ultimately the dark energy equation of state. Classical Cepheids are fundamental test bench for stellar physics,  optimal tracers of galactic structure, and more importantly they are primary distance indicators and historically the first rung of the local distance ladder that leads to  the Hubble Constant. The latest measurement using Cepheieds is from the S$\mbox{H}_0$ES team and yields the most precise among the late-time\footnote{For a recent review about late- and early-time measurements see \citet{verde19}.} values: $\mbox{H}_{0}=74.0\pm 1.4 \, \mbox{km s$^{-1}$ Mpc$^{-1}$}$ \citep{riess19}. This latter $~4\sigma$ is in tension with the Planck result, $\mbox{H}_{0}=67.4\pm 0.5 \, \mbox{km s$^{-1}$ Mpc$^{-1}$}$ \citep{planck18}, that is inferred from Cosmic Microwave Background, assuming $\Lambda\mbox{CDM}$. The tension increases over $5\sigma$ combining the independent late probes. Thus, while it becomes compelling to search for new physics beyond the standard cosmological model, reduce and control systematic errors on the late universe measurements is mandatory.

In this frame the S$\mbox{H}_0$ES value is the result of a huge effort to simplify the ladder and collect a larger and photometrically homogeneous sample, thus reducing dramatically the systematic errors along this ladder.
Nevertheless the classical ladder it is still affected by the uncertainties coming from the zero-point of the period-luminosity relation. One of these comes from the metallicity dependence of the relation that is still highly debated. 

From the theoretical point of view \citet{fiorentino13}, using non-linear convective pulsation models, predicted a metallicy effect of $\sim 0.4\, \mbox{mag/dex}$ with more metal rich Cepheids being fainter.  
Instead, \citet{anderson16} found an opposite trend, with more metal poor Cepheids being fainter, using linear non-adiabatic models. 

Observationally no appreciable effect was found by \citet{wielgorski17} and \citet{groenewegen18}. On the contrary \citet{romaniello08}, \citet{freedman11} and \citet{gieren18} found a photometric-band dependent effect. Nevertheless these studies found different magnitude of the effect and more notably a different sign. To be noted that only one study \citep{romaniello08} was conducted on direct Cepheids chemical abundances, although in a small sample, that limits the obtainable accuracy.

\section{The LACES project}

Locking the Abundances of CEpheids for SH$_0$ES (LACES) is a project aimed of assessing the metallicity effect on Leavitt law in  optical and NIR bands, through the direct spectrocopic measurements of chemical abundances over a large sample. Moreover it aims to link the Cepheid's abundance scale to the SNe~Ia host galaxies in
SH$_0$ES sample. The project is divided in three steps.

Magellanic Clouds Cepheids are a very suitable anchor since MCs geometric distance is very well known \citep{pietrzynski19} and their metallicities span a wide range $-1 \lesssim [\mbox{Fe}/\mbox{H}] \lesssim 0.0 \, \mbox{dex}$.
In these galaxies the LACES team collected spectra of more than 300 Cepheids with the FLAMES/GIRAFFE multi-object spectrograph at VLT. A sub-sample of these stars has been also observed with KMOS at VLT. To achieve $1\%$ error in distance, individual chemical abundances are required to have an error $<0.1 \, \mbox{dex}$.

The calibrated period-luminosity-metallicity relations will be used to calibrate the SNe~Ia maximum brightness in SNe~Ia host galaxies located at $\sim 40 \, \mbox{Mpc}$. Unfortunately, obtaining direct reliable chemical abundances for Cepheids at these distances is out of reach for current facilities, therefore Red Super Giants are used as metallicity proxy. An intermediate rung is thus needed to tie together Red Super Giant and Cepheid metallicity.

We concentrate here on the first rung of this chemical ladder, in particular how errors on temperature measurements can bias abundance measurements.

\section{Systematic abundance errors due to temperature measurements}

The degeneracy of the atmospheric a parameter can affect the chemical abundances measurements. The main one is the effective temperature, that can lead to a bias of~$\sim 0.1\, \mbox{dex}$ every $100 \,\mbox{K}$ of error, the sign being dependent on the true effective temperature. This is even more compelling for stars that can vary up to $1000 \, \mbox{K}$ their effective temperature during one pulsation cycle.

In order to avoid bias due to reddening, which affect color-temperature relations, we decided to implement the Line Depth Ratio (LDR) method, that relies on the sensitivity to temperature of metallic lines \citep{gray89}, empirically calibrated by \citet{proxauf18} using Galactic Cepheids, that encompass more than 200 line doublets in the wavelength coverage of the LACES optical spectra. 

To test how the estimated temperature is biased by the other atmospheric parameters and the instrumental set up, we built a synthetic spectra grid with Atlas9 (Kurucz) models. The grid spans the temperature range $3500-7800 \, \mbox{K}$, three surface gravity values: $log(\mbox{g})=0.0, \, 1.0, \, 2.0$, three metallicity values: $[\mbox{Fe}/\mbox{H}]=0.0, \, -0.5, \, -1.0$ and two spectral resolutions: $\mbox{R}=20000, \, 50000$.
The difference between the LDR temperature and the input temperature of the synthetic spectra, as a function of this latter, is shown on Figure~\ref{p18Synt}. These residuals can be as high as hundreds of Kelvin at the edge of the grid and also the gravity and resolution dependencies are  noticeable.


\articlefigure[scale=0.4]{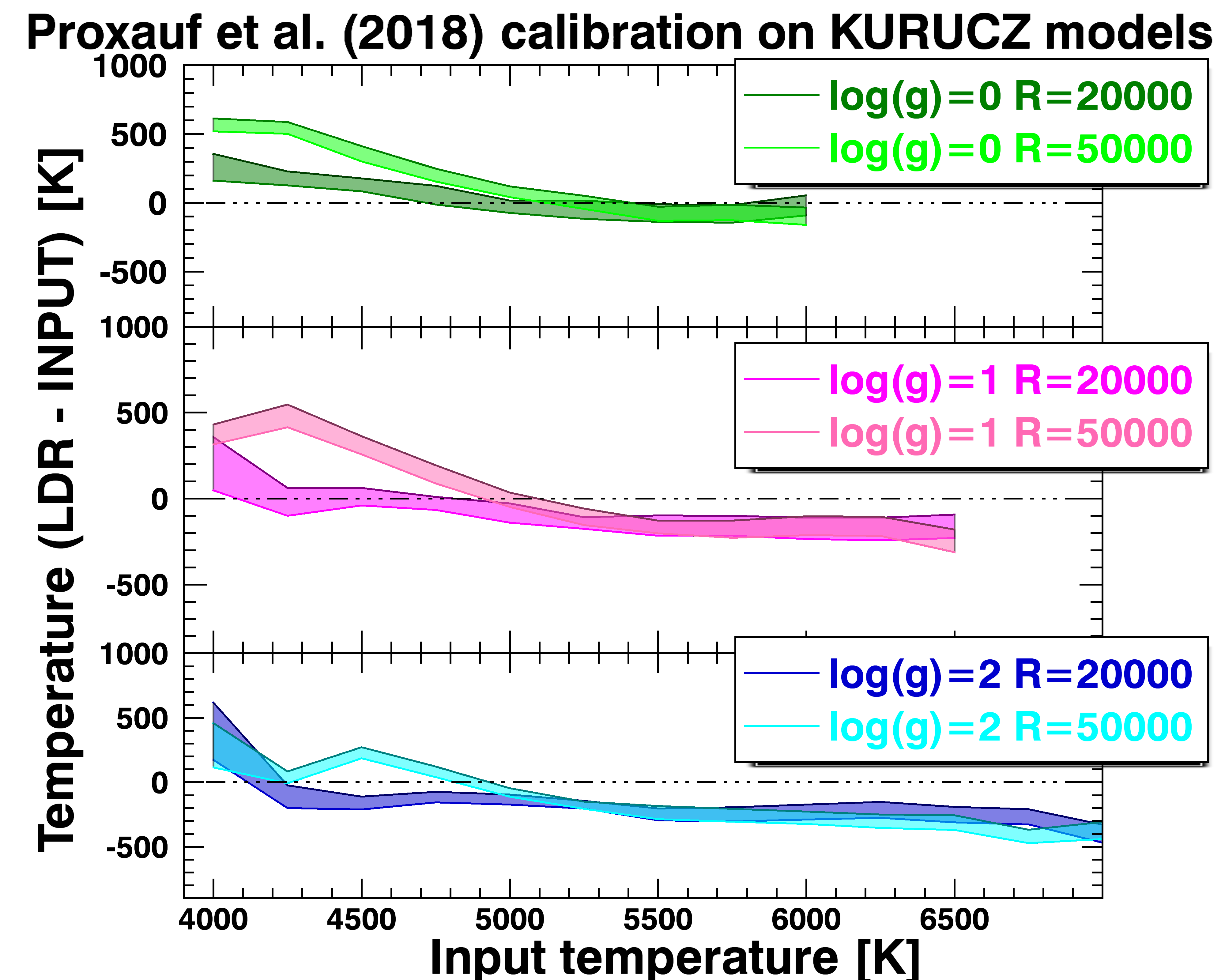}{p18Synt}{LDR temperature residuals applying \citet{proxauf18} calibration to Kurucz synthetic spectra at different surface gravity and resolution for $[\mbox{Fe}/\mbox{H}]=0.00$ (Mancino et al. in prep.).}

On Figure~\ref{pair65}, the relation between the input temperature and the measured line depth ratios is shown for an S~I~-~Ni~I doublet. For a measured depths ratio, different metallicity models yield different temperatures.
 
\articlefigure[scale=0.45]{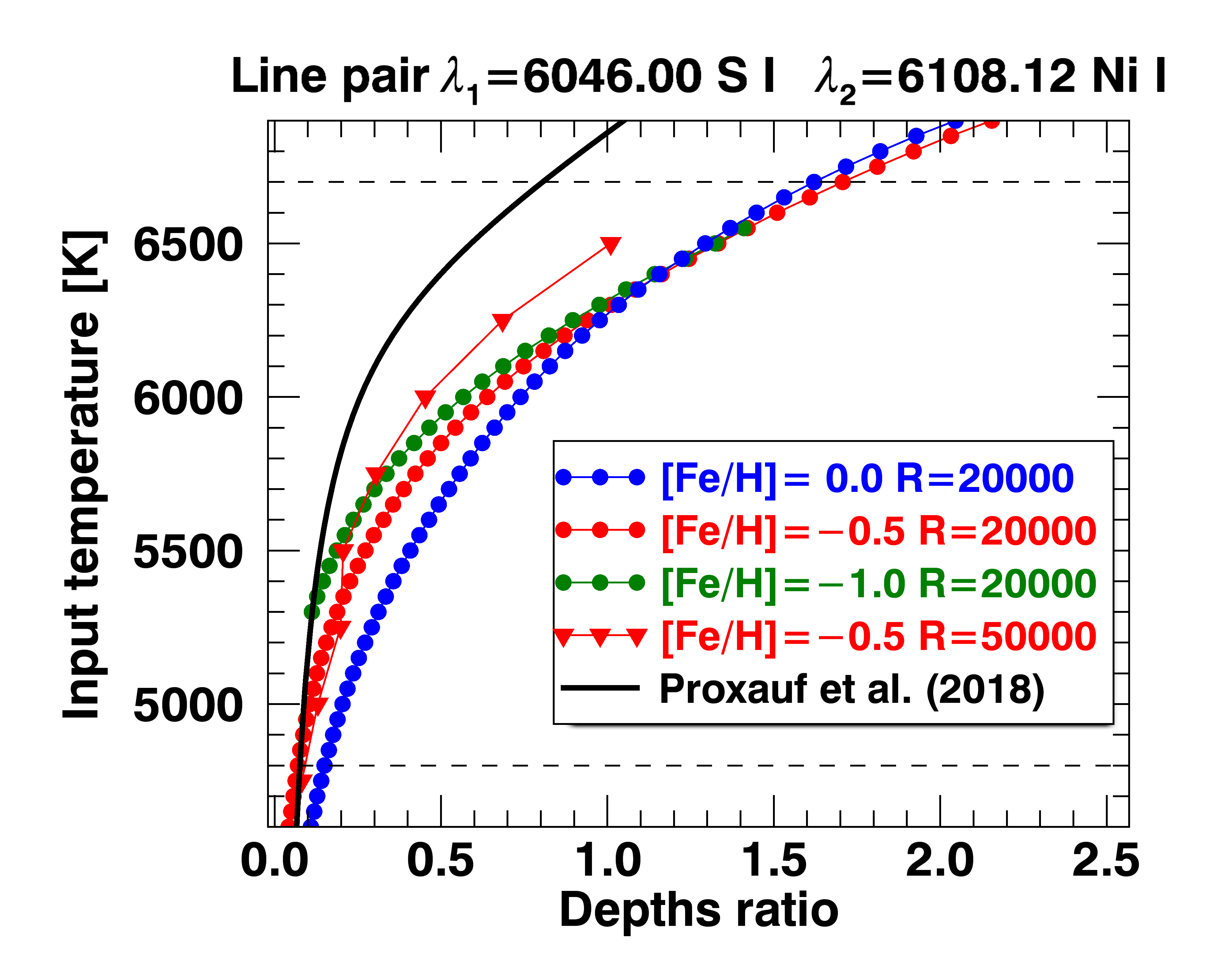}{pair65}{Line depth ratio for a line pair example measured for different set of synthetic spectra, varying iron content and resolution. The solid black line shows the reference calibration function. The horizontal-dashed lines mark the validity range for the specific doublet (Mancino et al. in prep.).}

To address this problem, we decided to calibrate the LDRs self-consistently over the synthetic spectra in order to minimize these dependencies. We used Chebyshev polynomials to fit temperature-ratio relations for each of the line pairs, minimizing the temperature residuals. We repeated the calibration over the three metallicity sets, and we applied them to the synthetic spectra grid to cross-check for metallicity dependence. This is shown in Figure~\ref{syntcalib}. The residuals are minimized by construction for the calibration at solar metallicity, but they show a trend with the input temperature when a different metallicity calibration is applied. At the colder edge of the grid the temperatures are overestimated of $\sim150\, \mbox{K}$ and $\gtrsim 300\, \mbox{K}$ applying respectively $[\mbox{Fe}/\mbox{H}]=-0.5$ and $[\mbox{Fe}/\mbox{H}]=-1.0$ LDR calibrations to solar metallicity spectra.

 This not only affects the chemical abundance measurements but it also changes the inferred properties of the star. For example, applying the LDR calibration made on different metallicity with respect to the observed Cepheid up to $\sim 3\%$ difference in mean temperature and up to $\sim20\%$ difference temperature amplitude.


\articlefigure[scale=0.40]{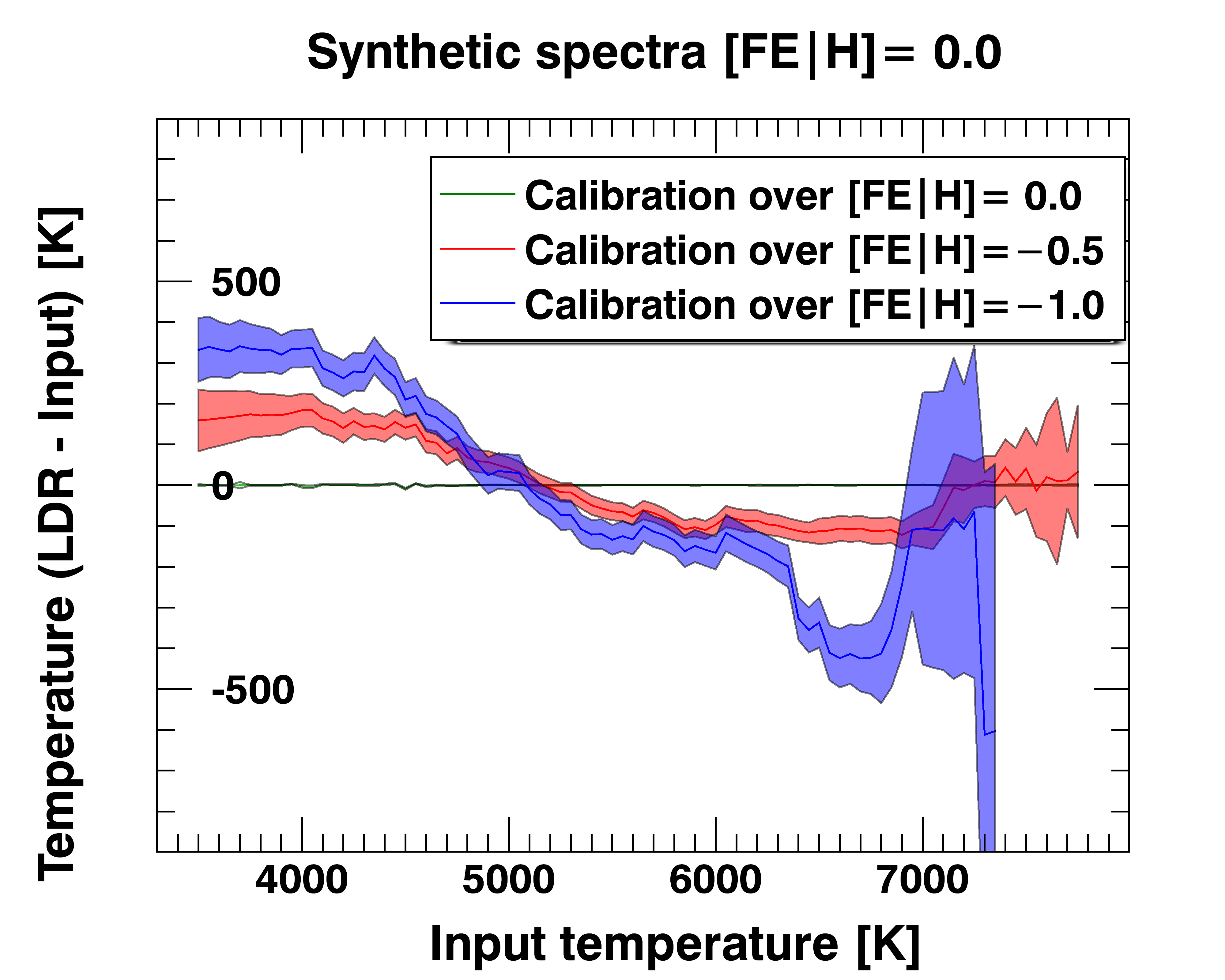}{syntcalib}{LDR temperature residual applying the calibration done over solar metallicity synthetic spectra to Kurucz models for the three metallicity values (Mancino et al. in prep.).}

We also analyzed the UVES spectra of LMC Cepheids, already studied by \citet{romaniello08}. In Figure~\ref{uves}, the temperatures measured applying the synthetic and empirical calibrations show striking differences. In order to minimize such effects, it is necessary to employ a calibration consistent with the models used for the abundance analysis.

\articlefigure[scale=0.45]{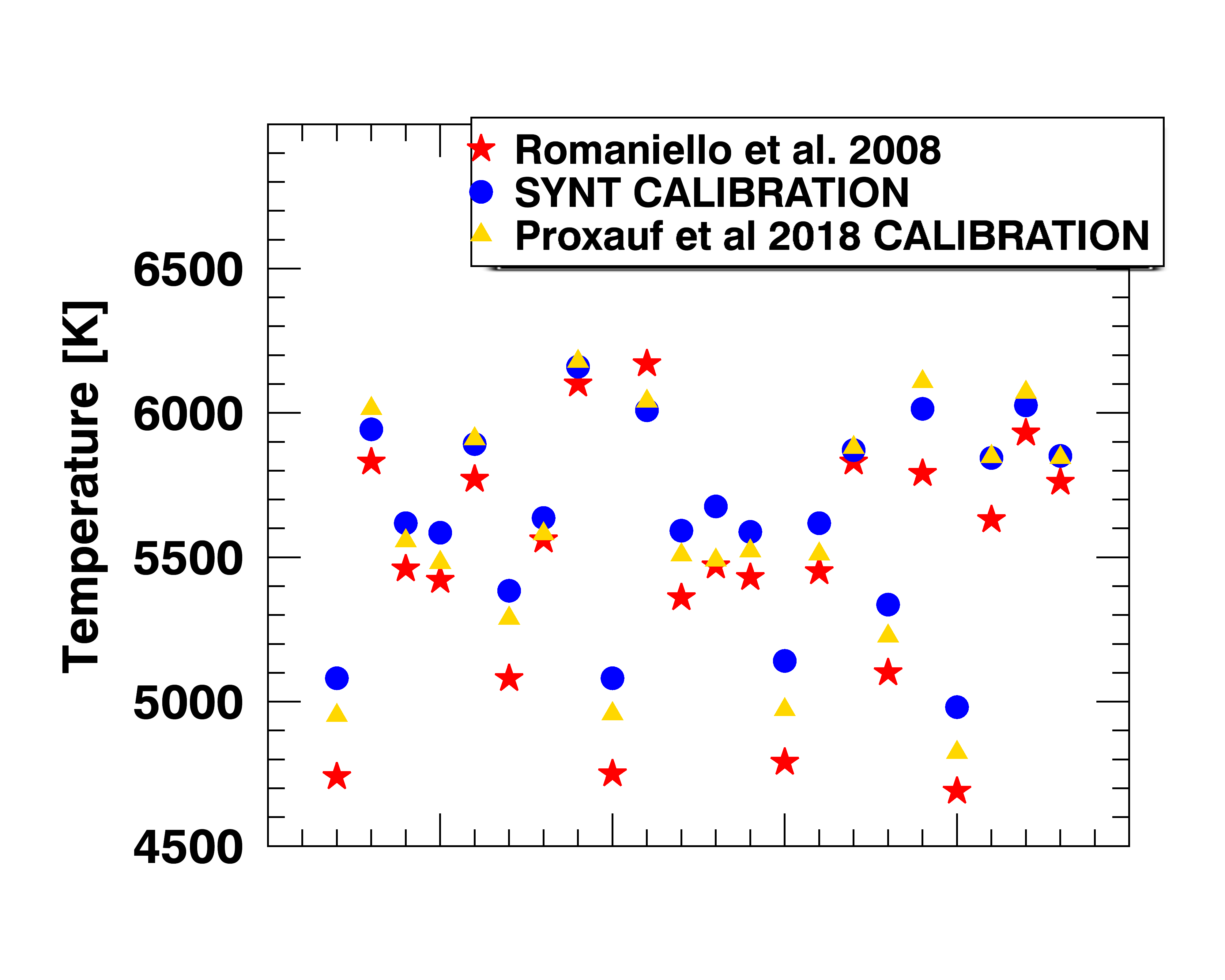}{uves}{LDR temperature for the LMC Cepheids from \citet{romaniello08} (Mancino et al. in prep.).}  
 
\section{Concluding remarks} 

Classical Cepheids have a primary role as distance indicators and they provide the most precise anchor for the local H$_0$ measurements. 
In order to  understand and solve the Hubble controversy, tight control of systematic errors is mandatory. In the classical distance ladder metallicity effects are not yet understood to the required level, the LACES team thus collected the largest spectroscopic sample of MCs Cepheids to have direct chemical abundance measurements. Through a synthetic spectra grid we investigated how systematic errors on LDR temperatures bias the abundance measurements although caveats on these measurements must be more carefully inquired.

\bibliography{BibliographySHORT} 

\end{document}